\documentclass[prd,english,preprintnumbers,amsmath,amssymb,nofootinbib,twocolumn,superscriptaddress]{revtex4-1}

\pdfoutput=1

\usepackage[latin1]{inputenc}
\usepackage{graphicx}
\usepackage{bbm}
\usepackage{amssymb}
\usepackage{amsmath}
\usepackage{tabularx}
\usepackage{color}
\usepackage{ulem}

\usepackage{dsfont}

\def\0#1#2{\frac{#1}{#2}}

\def\s0#1#2{\mbox{\small{$ \frac{#1}{#2} $}}}

\newcommand{\be}{\begin{eqnarray}}
\newcommand{\ee}{\end{eqnarray}}

\newcommand{\nn}{\nonumber }

\newcommand{\beq}{\begin{equation}}
\newcommand{\eeq}{\end{equation}}
\newcommand{\bea}{\begin{eqnarray}}
\newcommand{\eea}{\end{eqnarray}}
\newcommand{\Nc}{N_{\rm c}}

\newcommand{\dif}[1]{\ensuremath{\medspace \mbox{d} #1}}

\usepackage{babel}
\makeatother
\begin{document}

\title{Crystalline {Ground States} in 
Polyakov-loop extended Nambu--Jona-Lasinio Models}
\author{Jens Braun}
\affiliation{Institut f\"ur Kernphysik (Theoriezentrum), Technische Universit\"at Darmstadt, 
D-64289 Darmstadt, Germany}
\affiliation{ExtreMe Matter Institute EMMI, GSI, Planckstra{\ss}e 1, D-64291 Darmstadt, Germany}
\author{Felix Karbstein}
\affiliation{Helmholtz-Institut Jena, Fr\"obelstieg 3, D-07743 Jena, Germany}
\affiliation{Theoretisch-Physikalisches Institut, Friedrich-Schiller-Universit\"at Jena, Max-Wien-Platz 1, D-07743 Jena, Germany}
\author{Stefan Rechenberger}
\affiliation{Institut f\"ur Kernphysik (Theoriezentrum), Technische Universit\"at Darmstadt, 
D-64289 Darmstadt, Germany}
\author{Dietrich Roscher}
\affiliation{Institut f\"ur Kernphysik (Theoriezentrum), Technische Universit\"at Darmstadt, 
D-64289 Darmstadt, Germany}

\begin{abstract}
Nambu--Jona-Lasinio-type models have been used extensively to study the dynamics of 
the theory of the strong interaction at finite temperature and quark chemical potential on a phenomenological level. In addition to these studies, which
are often performed under the assumption that the ground state of the theory is homogeneous, searches 
for the existence of crystalline phases associated with inhomogeneous ground states 
have attracted a lot of interest in recent years. In this work, we study the Polyakov-loop extended Nambu--Jona-Lasinio model and find that the existence of a crystalline phase is stable against a variation of the parametrization of the underlying 
Polyakov loop potential. To this end, we adopt two prominent parametrizations. Moreover, we observe that
the existence of a quarkyonic phase depends crucially on the parametrization, in particular in the regime of the phase diagram
where inhomogeneous chiral condensation is favored.
\end{abstract}

\maketitle

%
\section{Introduction}
The original motivation to search for inhomogeneous, i.e. crystalline, ground states in many-body systems 
goes back to the early ground-breaking works by {\it Fulde} and {\it Ferrell} as well as {\it Larkin} and {\it Ovchinnikov}  
who found that the ground state of superconducting materials as described within {\it Bardeen-Cooper-Schrieffer} (BCS) theory~\cite{Bardeen:1957mv} 
is not necessarily homogeneous~\cite{FF,LO}. In recent years, the possibility of having inhomogeneous ground states in
a variety of different theories, such as the theory of the strong interaction, i.e. Quantum Chromodynamics (QCD),
has attracted a lot of attention, see, e.g., 
Refs.~\cite{Bringoltz:2006pz,Bringoltz:2009ym,Nickel:2009wj,Kojo:2009ha,Muller:2013pya,Buballa:2014tba,Carignano:2014jla,Lee:2015bva,Heinz:2015lua}.

The search for inhomogeneous phases in the phase diagram of QCD is inspired and paralleled by similar
searches in related condensed-matter systems (see, e.g., Refs.~\cite{Jackiw:1981wc,Mertsching,Machida,Chodos:1998wg,Kleinert:1998kj}) 
and ultracold atomic gases (see, e.g., Refs.~\cite{Bulgac:2008tm,Stoof,Radzihovsky}) 
where the emergence of inhomogeneous condensates can be related to the in-medium formation of bosonic bound states with finite
center of mass momentum~\cite{Roscher:2013cma,Roscher:2015xha}.
However, studies of -- at least within the mean-field approximation -- exactly solvable, relativistic field theories of strongly interacting fermions in low dimensions probably played the most
important role in recent years~\cite{Schon:2000qy,Thies:2003br,Schnetz:2004vr,Schnetz:2005ih,Basar:2008im,Basar:2008ki,Basar:2009fg}.
The analyses of these low-dimensional models
are most relevant to guide our understanding of
fermionic theories in higher dimensions. Three-dimensional Nambu--Jona-Lasinio-type (NJL) models are 
close relatives of these low-dimensional models and are often employed as low-energy
models for QCD. In particular, they are used to study the finite-temperature phase structure of QCD on a phenomenological level,
see Refs.~\cite{Klevansky:1992qe,Buballa:2003qv,Fukushima:2011jc} for reviews. In recent years, studies of these models have been extended
in various ways and the search for inhomogeneous phases has been put forward~\cite{Nickel:2009wj,Kojo:2009ha}. In fact, 
depending on the actual choice for the parameters, an inhomogeneous phase has been found to exist at large quark chemical
potential in this class of models~\cite{Nickel:2009wj,Kojo:2009ha}, see Ref.~\cite{Buballa:2014tba} for a recent review. 
This exotic phase even persists to exist when the NJL model is set up with a 
nonlocal four-fermion interaction~\cite{Carlomagno:2014hoa,Carlomagno:2015nsa}.

In the past twenty years several extensions of the original NJL model \cite{Nambu:1961tp,Nambu:1961fr} have been introduced to bring it closer to QCD. Some focus
has been laid on extensions which account at least partially for the missing dynamics of the gluon degrees of freedom.
The so-called Polyakov-loop extended NJL (PNJL) model played a prominent role in this regard~\cite{Meisinger:1995ih,Fukushima:2003fw,Ratti:2005jh}
and also inspired the discussion on the existence of a confined but chirally symmetric phase, the quarkyonic phase~\cite{McLerran:2007qj}.
Within this extension, the order parameter for the deconfinement phase transition, the Polyakov loop, is coupled to the chiral quark
dynamics as described by the NJL model and therefore it allows to study to some extent the effect of the confining gauge dynamics
on the chiral phase structure of QCD. By now, first studies are available which aim at an
understanding of the effect of the gauge dynamics on the chiral dynamics driving the formation of an inhomogeneous ground state
at large quark chemical potential, see, e.g. \cite{Partyka:2010em,Carignano:2010ac}, and Ref.~\cite{Buballa:2014tba} for a review. 
Unfortunately, the parametrization of the Polyakov loop potential
underlying PNJL models is ambiguous. 
In the original formulations of PNJL-type models, 
the backreaction of the quark dynamics on the Polyakov loop potential has not been taken into account, see Ref.~\cite{Haas:2013qwp} for a discussion
of such effects. For example, the Polyakov loop potential in full QCD has an implicit dependence on 
the quark chemical potential and the number of quark flavors. The latter can readily be understood by recalling the
dependence of the dynamically generated scale of QCD, namely~$\Lambda_{\text{QCD}}$, on the number of 
quark flavors. This quantity directly affects the scaling of all (dimensionful) physical observables, 
such as the chiral phase transition temperature~\cite{Braun:2005uj,Braun:2006jd}.
This has also been studied in Ref.~\cite{Schaefer:2007pw}, where the dependence on the quark chemical
potential has been considered as well, and used to amend Polyakov loop potentials entering QCD low-energy models. 
Taking this into account, it has been found that the chiral and deconfinement phase transition lines remain close to each other up
to large values of the quark chemical potential, implying the disappearance of the above-mentioned 
quarkyonic phase in these model studies~\cite{Schaefer:2007pw}.

Since the parametrization of the Polyakov loop potential affects the phase structure of these QCD models, 
particularly at large quark chemical potential \cite{McLerran:2007qj,Schaefer:2007pw},
it is also important to study the fate of inhomogeneous phases under a variation of this parametrization. In this paper,
we address this question. In Sect.~\ref{sec:model}, we introduce our model setup.
To detect the onset of inhomogeneous phases, we employ the {\it fermion doubling trick} introduced in Ref.~\cite{Braun:2014fga}. This approach 
has proven to reproduce the correct phase structure of the one-dimensional 
Gross-Neveu (GN) model in the mean-field approximation~\cite{Thies:2003kk,Schnetz:2004vr,Thies:2006ti}, including 
the existence of the inhomogeneous phase.
Beyond the phenomenological 
questions being at the heart of this work, our present study also represents a further demonstration of the applicability of 
this trick~\cite{Braun:2014fga}.
In Sect.~\ref{sec:ps}, we present our results for the phase structure of the PNJL model 
and discuss the effect of the parametrization of the Polyakov loop
potential. Our conclusions are given in Sect.~\ref{sec:conc}.

\section{Model}\label{sec:model}
In the following we work in four-dimensional Euclidean spacetime. 
Focusing exclusively on the chiral limit, i.e. the limit of vanishing current quark 
masses, the action of the (local) PNJL model is given by~\cite{Fukushima:2003fw,Ratti:2005jh}
\be
&& S =  \int_{\tau}\int_{\vec{x}}\Big\{\mathcal{U}\left(\Phi,\bar{\Phi},T\right) + \bar{\psi}\left({\rm i}\gamma_\nu\left(\partial_\nu \!-\! {\rm i}\bar{g}A_0\delta_{\nu,0}\right) \!-\! {\rm i}\mu\gamma_0\right)\psi \nn\\
&& \qquad\qquad\qquad\qquad\qquad +\, \frac{\bar{\lambda}}{2}\left[\left(\bar{\psi}\psi\right)^2 \!+\! \left(\bar{\psi} {\rm i}\gamma_5 \vec{t}\, \psi\right)^2\right] \Big\}\,,
\label{eq:S}
\ee
where $\cal U$ denotes the Polyakov loop potential, $\mu$ is the quark chemical potential, the $t_i$'s are the Pauli matrices,
$\tau$ is the Euclidean time, $\int_{\tau}=\int_0^{\beta} \dif\tau$, and $\int_{\vec{x}} = \int \dif^3x$ accordingly, and finally $\beta=1/T$ is the inverse temperature.
The quark fields~$\psi$ interact via a local four-fermion interaction parametrized by the bare coupling~$\bar{\lambda}$. Moreover, we have introduced a temporal gauge field~$A_0$
which can be related to the order parameter for the deconfinement phase transition, 
namely the expectation value of the Polyakov loop~$\langle \Phi(\vec{x}) \rangle$. The corresponding variables~$\Phi(\vec{x})$ and $\bar\Phi(\vec{x})$
read~\cite{Polyakov:1978vu,Susskind:1979up}
\be
\Phi(\vec{x}) = \frac{1}{\Nc}\mbox{Tr}_{\rm c}L(\vec{x})\,,\quad {\bar\Phi(\vec{x}) = \frac{1}{\Nc}\mbox{Tr}_{\rm c}L^\dag(\vec{x})\,,}
\ee
with
\be
L(\vec{x})=\mathcal{P}\,\mbox{exp}\biggl({\rm i}\bar{g}\int_0^\beta\!\dif{\tau}A_0(\tau,\vec{x})\biggr)\,.
\ee
Here, $ \mathcal{P}$ denotes path ordering. The color trace, $\mbox{Tr}_{\rm c}$, is taken in the fundamental representation.
The quantity $\bar{g}$ is the bare gauge coupling, 
which can be conveniently absorbed into a redefinition of the gauge field, and $\Nc=3$ is the number of colors.

The expectation value $\langle \Phi(\vec{x}) \rangle$ measures whether center symmetry is realized in the ground state of the theory~\cite{Svetitsky:1985ye,Greensite:2003bk}.
A center-symmetric confining ground state is signaled by $\langle\Phi\rangle=0$, whereas deconfinement, $\langle\Phi\rangle\neq 0$, 
is associated with center-symmetry breaking. Moreover, the negative logarithm of~$\langle\Phi\rangle$ 
relates to the free energy of a static fundamental color source~\cite{Svetitsky:1985ye,Greensite:2003bk}. 
In the presence of (light) dynamical quarks as in this work, the center symmetry
is broken explicitly, i.e. $\langle\Phi\rangle \neq 0$ for all temperatures, rendering the deconfinement phase transition a crossover. Note that~$\Phi$ 
and~$\bar{\Phi}$ can be considered as independent fields. The expectation values of both are identical at $\mu=0$ but
differ at finite~$\mu$, $\langle\bar{\Phi}\rangle \geq \langle\Phi\rangle$~\cite{Dumitru:2005ng}.

In our present study, the quark dynamics is coupled to the Polyakov 
loop variables~$\Phi$ and $\bar\Phi$ via the temporal gauge field~$A_0$. Their dynamics is determined
by the potential~$\mathcal U$ in the action~\eqref{eq:S}. In pure gauge theory, 
the potential~$\mathcal U$ may be used to define an effective theory for the Polyakov loop dynamics 
in terms of the independent fields~$\Phi$ and~$\bar{\Phi}$.
For~$\mathcal U$, we choose the following ansatz which respects 
the $Z(3)$ center symmetry of the underlying $SU(3)$ gauge theory~\cite{Pisarski:2000eq,Pisarski:2006hz}:
\be
\frac{\mathcal{U}}{T^4} = -\frac{b_2}{2}\bar{\Phi}\Phi - \frac{b_3}{6}\left(\Phi^3 + \bar{\Phi}^3\right) + \frac{b_4}{4}(\bar{\Phi}\Phi)^2\,.\label{eq:U}
\ee
Restricting ourselves to constant and homogeneous $\Phi$ and $\bar\Phi$, and dropping fluctuations, the potential~$\mathcal U$
can be directly related to the associated effective action. 
Its minimization then yields the temperature-dependent
expectation values $\langle\Phi\rangle$ and~$\langle\bar{\Phi}\rangle$
defining the minimum of the potential as well as the thermodynamic quantities, such as the pressure $p(T)=-\mathcal{U}\left(\langle\Phi\rangle,\langle\bar{\Phi}\rangle,T\right)$.
This is used to fit the parameters in Eq.~\eqref{eq:U} to data
from lattice Monte Carlo simulations of $SU(3)$ Yang-Mills theory such that   
the behavior of thermodynamic quantities is 
consistent with these simulations and $\langle \Phi\rangle = \langle \bar{\Phi}\rangle$, as it should be. Here, we follow Refs.~\cite{Ratti:2005jh,Schaefer:2007pw} and employ
\be
&& b_2(T) = a_0 + a_1\left(\frac{T}{T_0}\right) + a_2\left(\frac{T}{T_0}\right)^2 + a_3\left(\frac{T}{T_0}\right)^3\,,\nn\\
&&\qquad\qquad\quad\; b_3 = 0.75\,,\quad b_4 = 7.5\,,
\ee
where
\be
&& a_0 = 6.75\,,\; a_1 = -1.95\,,\; a_2 = 2.625\,,\; a_3=-7.44\,.
\ee
The value of $T_0$ may be assumed to depend on the number of quark flavors and the quark chemical potential. This then accounts for the
fact that the dynamics of the gauge fields effectively generating the potential~$\mathcal U$ is in general affected by the quark dynamics. 
Following Ref.~\cite{Schaefer:2007pw},
we use~$T_0=208\,\text{MeV}$ at~$\mu=0$ for two quark flavors. Moreover, we consider two different parametrizations which are distinguished by either choosing
$T_0$ to be independent of $\mu$ or to be $\mu$-dependent as put forward in Ref.~\cite{Schaefer:2007pw}:
\be
T_0(\mu)=T_{\tau}{\rm e}^{-\frac{1}{\alpha_0 b(\mu)}}\,,\label{eq:t0}
\ee
where, for two massless quark flavors,
\be
b(\mu)=\frac{29}{6\pi}-{\frac{32}{\pi}}\left(\frac{\mu}{T_{\tau}}\right)^2\,,
\ee
and
\be
\alpha_0=0.304\,,\; T_{\tau}=1770\,\text{MeV}\,.
\ee
These parameters basically reflect the fact that the gauge coupling is 
assumed to be fixed at the $\tau$-mass scale~$m_{\tau} \approx T_{\tau}$, $\alpha_0 \approx \alpha(m_{\tau})$.

Let us now turn to the computation of the effective potential from the action~\eqref{eq:S}. To this end, we resort to
the mean-field approximation and perform the conventional Hubbard-Stratonovich transformation
of the action~\eqref{eq:S} which introduces auxiliary bosonic 
fields into the action (see, e.g., Refs.~\cite{Klevansky:1992qe,Buballa:2003qv,Braun:2011pp}),
\be
\sigma \sim (\bar{\psi}\psi)\,,\quad \vec{\pi} \sim (\bar{\psi}{\rm i}\gamma_5\vec{t}\,\psi)\,.
\ee
From a phenomenological point of view, these fields may be associated with the $\sigma$ and pion fields, respectively. 
In our search for inhomogeneous ground states,
we restrict ourselves to one of the most simple inhomogeneous ans\"atze for the order parameter fields,
\be
\sigma(\vec{x}) = \bar{\sigma}\cos\bigl(2\vec{Q}\cdot\vec{x}\bigr)\,,\quad{\vec{\pi}(\vec{x})=0\,,}\label{eq:scos}
\ee
with amplitude $\bar{\sigma}$ and wave-vector $\vec{Q}$.
Another simple and prominent ansatz is the complex chiral density wave, cf., e.g. \cite{Sadzikowski:2000ap,Nakano:2004cd}. 
For the latter, the (mean field) effective potential may even be computed exactly. 
However, it comes along with a generally nonvanishing parity-breaking order parameter $\langle\vec{\pi}(\vec{x})\rangle$.

For~$Q\equiv|\vec{Q}|=0$, Eq.~\eqref{eq:scos} reduces to the ansatz conventionally
used in studies where the possibility of inhomogeneous chiral condensation is not taken into account. 
Switching to momentum space and employing the fermion doubling trick~\cite{Braun:2014fga},
we assume that the quark fields $\psi_n(\vec{p}-\vec{Q})$ and $\psi_n(\vec{p}+\vec{Q})$ are independent degrees of freedom 
in the Hubbard-Stratonovich-transformed action and integrate them out straightforwardly, see also Sect.~\ref{sec:ps} for a discussion of 
the limitations coming along with this assumption. 

Choosing the temporal gauge field~$A_0$ to be in the Cartan subalgebra and time- as well as 
space-independent~\cite{Meisinger:1995ih,Fukushima:2003fw,Ratti:2005jh}, we obtain the following result for the effective potential~$\Omega$:
\begin{widetext}
\be
\Omega(\Phi,\bar{\Phi},\bar{\sigma},T,\mu,{Q})
&=& \mathcal{U}(\Phi,\bar{\Phi},T) + \frac{T}{V}\int_\tau\int_{\vec{x}}\frac{\sigma(\vec{x})^2}{2\bar{\lambda}} - 2 T\int_{\vec{p}} \sum_{\pm}\left\{ \ln\left[1+3\left(\Phi+\bar{\Phi}e^{\left(\mu-E_\pm\right)/T}\right)e^{\left(\mu-E_\pm\right)/T} + e^{3\left(\mu-E_\pm\right)/T}\right] \right.\nonumber\\
&&\left. \qquad +\ln\left[1+3\left(\bar{\Phi}+\Phi e^{-\left(\mu+E_\pm\right)/T}\right)e^{-\left(\mu+E_\pm\right)/T} + e^{3\left(\mu+E_\pm\right)/T}\right] \right\} - 6\int_{\vec{p}} \big\{E_++E_-\big\} 
\label{eq:ep}
\ee
\end{widetext}
with~$V$ being the spatial volume and
\be
E_{\pm}=\sqrt{\vec{p}^{\,2}+\vec{Q}^{2}+\bar{\sigma}^2\pm 2\sqrt{(\vec{p}\cdot\vec{Q})^2+\bar{\sigma}^2{\vec{Q}^2}}}\,.
\ee
Moreover, we choose a sharp ultraviolet (UV) 
cutoff~$\Lambda$ to regularize the space-like momentum integrals, i.e. $\int_{\vec{p}}:=\int_{|\vec{p}|<\Lambda}\frac{d^3p}{(2\pi)^3}$. Note that
we regularize both the zero- and finite-temperature contributions in Eq.~\eqref{eq:ep}.\footnote{In principle, it is possible to regularize only the zero-temperature 
contributions as the finite-temperature contributions are finite anyways. We have checked both types of regularization and found that, e.g., the chiral phase transition temperature at~$\mu=0$
increases by about $20\,\text{MeV}$ when both the zero- and finite-temperature contributions are regularized consistently with our sharp-cutoff prescription.}
\begin{figure}[t]
\includegraphics{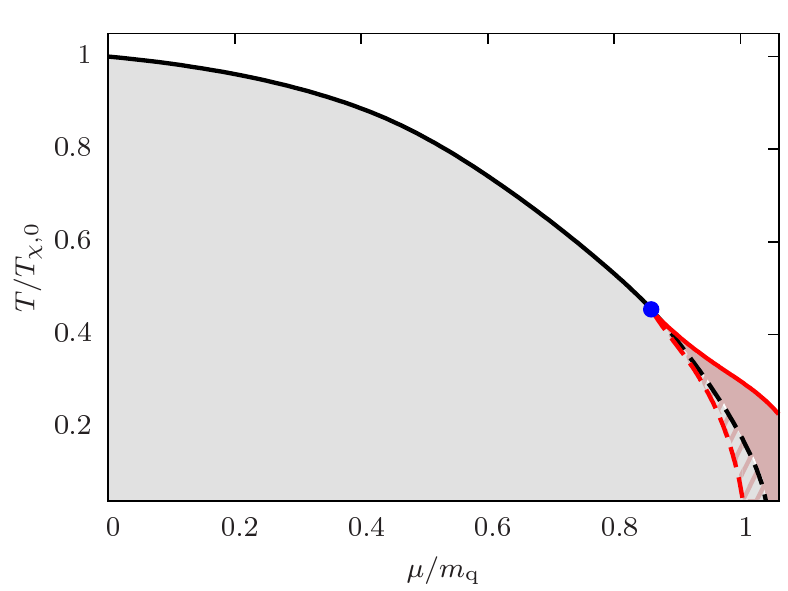}
\caption{(color online) Chiral phase diagram of the PNJL model with two massless quark flavors in the plane spanned by the temperature in units of $T_{\chi,0}\equiv T_{\chi}(\mu\!=\!0)$
and the quark chemical potential in units of {$m_q$} obtained with a $\mu$-dependent $T_0$-parameter.
The black solid/dashed line denotes the $2$nd/$1$st-order transition for the homogeneous setup, i.e., setting $Q\equiv0$ in Eq.~\eqref{eq:scos}. 
Both lines meet at a critical point (blue dot). The energetically favored inhomogeneous chiral condensate is found between the red solid and red dashed line, 
being of $2$nd and $1$st order respectively.}
\label{fig:pd}
\end{figure}

Similar to conventional (homogeneous) PNJL model studies~\cite{Ratti:2005jh,Schaefer:2007pw},
the ground state for a given temperature and chemical potential is then determined by searching numerically for the saddle point in the multidimensional space spanned by the 
quantities $\Phi$, $\bar{\Phi}$, $\bar{\sigma}$, and $Q$. The coupling constant~$\bar{\lambda}$ and the UV cutoff~$\Lambda$ are tuned to obtain
the constituent quark mass $m_q\approx\bar{\sigma}_0=325\,\text{MeV}$ and the pion decay constant $f_\pi = 92.4\,\text{MeV}$ independently of $Q$ at $T=\mu=0$.
Here, $\bar{\sigma}_0$ denotes the value of~$\bar{\sigma}$
minimizing the effective potential. To be specific, in accordance with Ref.~\cite{Ratti:2005jh}, this is achieved by 
choosing $\bar{\lambda}\Lambda^2\approx 4.34$ and~$\Lambda=651\,\text{MeV}$.
\begin{figure*}[t]
\includegraphics{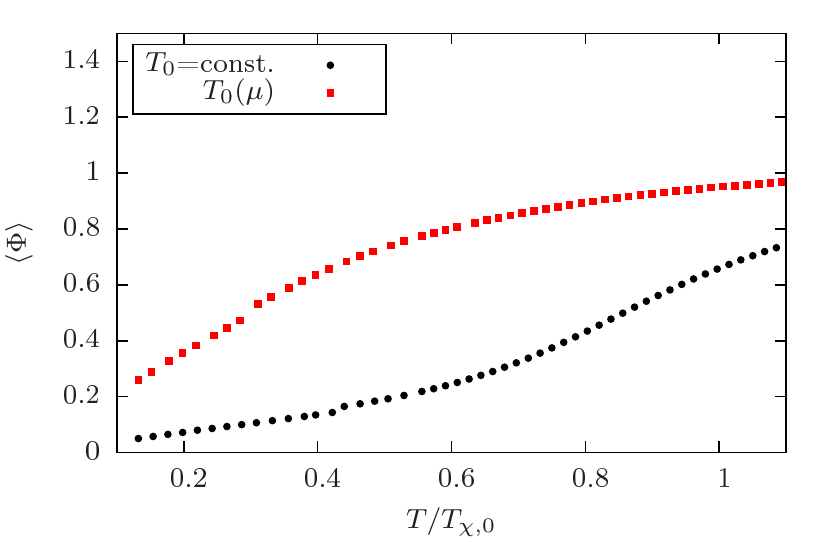}
\includegraphics{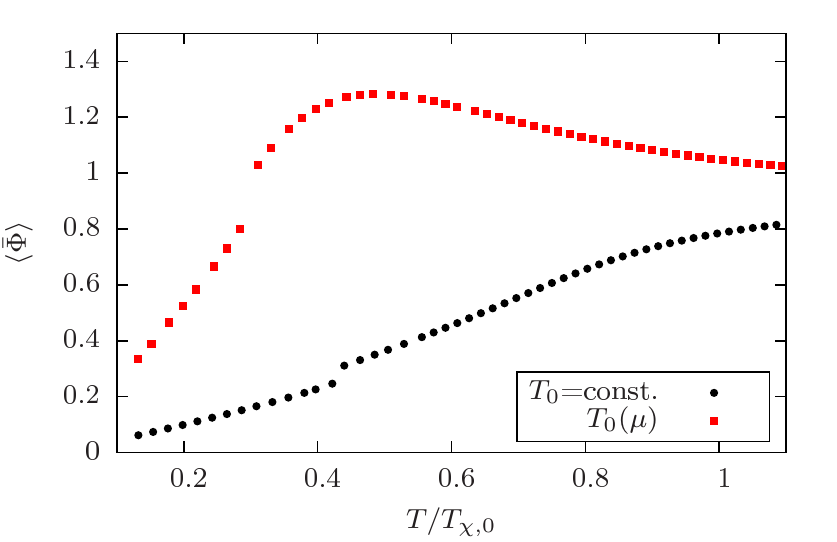}
\caption{(color online) Polyakov loop (left panel) and adjoint Polyakov loop (right panel) as a 
function of the temperature in units of~{$T_{\chi,0}$}
for the specific choice of $\mu/m_q=1.02$ for both $T_0=208\,\text{MeV}$ and a $\mu$-dependent 
parametrization of $T_0$, see main text for details.}
\label{fig:ploops}
\end{figure*}
\section{Phase Structure}\label{sec:ps}
\subsection{Chiral Phase Structure}

Let us now discuss the phase structure of the PNJL model as obtained from the two different 
parameterizations of the Polyakov loop potential in terms of the quantity~$T_0$, i.e.~$T_0=\text{const.}$
and the $\mu$-dependent ansatz~\eqref{eq:t0} for $T_0$, respectively.

In Fig.~\ref{fig:pd} we show our result for the chiral phase diagram of the PNJL model 
for the parametrization~\eqref{eq:t0} in the plane spanned by the temperature measured in units of
the chiral phase transition temperature at~$\mu=0$,~$T_{\chi,0}\equiv T_{\chi}(\mu{=0})$, and the quark
chemical potential~$\mu$ in units of the constituent quark mass {$m_q$}. 
For~$\mu=0$, we find~$T_{\chi}=227\,\text{MeV}$. Note that, as discussed in the previous section, this value depends on the
regularization procedure for a given set of parameters in the presently used mean-field approximation. Since we are primarily interested
in a qualitative discussion of the effect of the gauge dynamics as parametrized in terms of the Polyakov loop potential on the chiral phase structure at large
chemical potential, we do not tune this value any further here although it is somewhat larger than na\"ively expected for QCD with two massless quark flavors.

The chiral phase structure as obtained by employing a constant~$T_0$ agrees qualitatively with the one shown in Fig.~\ref{fig:pd}
which has been computed by using the ansatz~\eqref{eq:t0} for $T_0$. However,
there are differences in the behavior of the Polyakov loop in particular at large values of the chemical potential to be analyzed below. 

Let us start by discussing how the chiral phase structure is altered when we allow for the existence of inhomogeneous phases.
Not treating~$Q$ in Eq.~\eqref{eq:ep} as a variational parameter in our calculation but setting it to zero, we recover the 
conventional homogeneous chiral phase structure of the PNJL model, see, e.g., Ref.~\cite{Fukushima:2011jc} for a review. 
To be more specific, in Fig.~\ref{fig:pd}, the gray-shaded area bounded by the black
solid (second-order chiral phase transition) and black dashed (first-order chiral phase transition) line
is associated with the spontaneous formation of a homogeneous chiral condensate. 
The white area depicts the chirally symmetric phase.

Considering now $Q$ as a variational parameter, we observe the emergence of an inhomogeneous phase associated with 
a finite spatially varying order parameter at large quark chemical potential. 
This regime is bounded by the red dashed (first-order transition\footnote{Note that this first-order transition between
the phase with a homogeneous chiral condensate and the inhomogeneous phase is not related to chiral symmetry breaking
but to translation symmetry breaking as  measured by the value of~$Q$ in the ground state.}) and red solid (second-order chiral transition) line. 
Here, we have~$\bar{\sigma} >0$ and~$Q > 0$ in the ground state of the effective potential~\eqref{eq:ep}.
Notably, the first-order transition line is shifted to smaller values of the chemical potential 
as compared to the $Q=0$ case, implying that the size of the regime governed by a homogeneous
chiral condensate (gray-shaded area) shrinks. Note that
we only show the phase diagram in the regime~$\mu/m_q \lesssim 1.1$. For larger values of~$\mu$, the results suffer strongly by
the presence of the finite UV cutoff~$\Lambda$ which also constrains the domain of validity of the PNJL model as a low-energy 
model for QCD, independent of the details
of the chosen regularization scheme (e.g. sharp cutoff or exponential cutoff). In any case, the general features of the chiral phase 
structure observed in our present work are in accordance with previous studies, see Ref.~\cite{Buballa:2014tba} for a review.

Before we analyze the effect of the two different parametrizations of the Polyakov loop potential, let us discuss the strengths and the limitations of our
present approach. First of all, we work here in the mean-field approximation. Thus, fluctuation effects are not taken into account but they are certainly 
relevant to clarify conclusively whether an inhomogeneous phase is energetically favored at large chemical potential. To address this issue, it may in the future
be worthwhile to study, e.g., the renormalization group (RG) flow of the auxiliary-field propagators as recently done in the context of ultracold Fermi gases to
detect the onset of inhomogeneous condensation~\cite{Roscher:2015xha}. Second, our search for inhomogeneous phases is based on the fermion doubling trick
\cite{Braun:2014fga} which allows to detect exactly the position of general second-order chiral phase transitions but
only allows for an approximate determination of transition lines between phases with a homogeneous and an inhomogeneous chiral condensate (red dashed line
in Fig.~\ref{fig:pd}). In the context of the one-dimensional GN model, however,  it has been shown numerically and argued analytically that at least the location
of this type of phase transition line is also reproduced reasonably well.

Finally, we have restricted ourselves to a simple single-cosine
ansatz~\eqref{eq:scos} for the inhomogeneous ground state in the $\sigma$ channel. 
Again, for the GN model in one dimension,
it has been shown that the exact ground-state solution approaches this single-cosine form 
close the second-order phase boundary~\cite{Thies:2003kk,Schnetz:2004vr,Thies:2006ti}. Although the predictive power of 
the ansatz~\eqref{eq:scos} has been demonstrated for the one-dimensional GN model~\cite{Braun:2014fga}, and also for one-dimensional
non-relativistic field theories~\cite{Roscher:2013cma}, the exact functional form of the 
energetically most favorable inhomogeneity in higher dimensional GN- or NJL-type
models is still unknown. In fact,
there is a plethora of possible functional forms for the inhomogeneous ground state beyond our present one-dimensional ansatz~\eqref{eq:scos} in the $\sigma$ channel, 
ranging from so-called chiral spirals~\cite{Kojo:2009ha} to higher dimensional inhomogeneities~\cite{Buballa:2014tba}.

In this respect, we add that, in our study with $\mu$-dependent $T_0$, we find 
a very thin spike of the inhomogeneous phase (width~$\ll 1\,\text{MeV}$ in $T$-direction)
which `cuts' into the phase with a homogeneous chiral condensate starting from the so-called 
{\it Lifshitz} point (blue dot in Fig.~\ref{fig:pd}). This may be considered as an indication that
our ansatz for the inhomogeneity is not sufficient to determine correctly the location of the first-order transition line (red dashed line). 
On the other hand, it could very well 
be an artifact generated by the specific parametrization used here for the Polyakov loop potential or even a numerical artifact.
Indeed, we have not observed indications for this in our studies with constant~$T_0$ as well
as in our benchmark studies of the conventional NJL model.

In any case, based on the ansatz~\eqref{eq:scos}, our present approach allows to detect regimes of inhomogeneous chiral condensation in the phase diagram in
a numerically inexpensive way as it only relies on the evaluation of the effective potential~$\Omega$, as also usually 
done in conventional PNJL model studies only allowing for homogeneous chiral condensation. 
In particular, our approach allows to study efficiently the interplay of inhomogeneous 
condensation and confining dynamics as modelled by the Polyakov loop potential and can therefore help to guide numerically more costly approaches, 
such as exact diagonalization methods.

\subsection{Effect of Confining Dynamics}

Let us now discuss the effect of the parametrization of the Polyakov loop potential 
on the phase structure, with an emphasis
on the regime at large chemical potential. Using a $\mu$-independent constant value for~$T_0$, it has been found that
a quarkyonic phase emerges where the ground state is confined but chirally symmetric~\cite{McLerran:2007qj}. On the other
hand, it has been observed that this phase diminishes or even vanishes completely when~$T_0$ is assumed to be~$\mu$-dependent as given
in Eq.~\eqref{eq:t0}, see Ref.~\cite{Schaefer:2007pw}.  
In our search for inhomogeneous phases, we have used these two prominent and frequently used choices for~$T_0$. As already mentioned above,
we observe that the chiral phase structure does not depend on our choice for~$T_0$ on a qualitative level. Quantitatively, we find
that the {\it Lifshitz} point is shifted to lower temperatures and larger values of the quark chemical potential when we choose~$T_0$ to be $\mu$-dependent.
As a direct consequence, we observe that the inhomogeneous phase tends to shrink in this case. 

We now turn to the deconfinement order parameter. Whereas our results for the deconfinement and the chiral phase transition line are in accordance
with previous results in the regime to the left of the critical point~\cite{McLerran:2007qj,Schaefer:2007pw,Fukushima:2011jc}, 
they differ to the right of this point. Setting~$T_0=\text{const.}$, we 
find that, for a given fixed value of~$\mu$, 
the deconfinement phase transition as measured in terms of~$\langle \Phi\rangle$ (and~$\langle \bar{\Phi}\rangle$) occurs at temperatures
well above the chiral phase transition separating the inhomogeneous phase and the chirally symmetric phase, see Fig.~\ref{fig:ploops}.
This suggests the existence of a confined phase with restored chiral symmetry~\cite{Buballa:2014tba}, i.e. a quarkyonic phase. 
A comment is in order here. Since we work with light dynamical quarks, the deconfinement transition is a crossover rather than a true phase transition. Thus, the definition of the actual
phase transition temperature is to some extent ambiguous. For example, we may use~$\langle \Phi\rangle(T_{\rm d})=0.5$ to define
the deconfinement temperature~$T_{\rm d}$. In any case, we have only
$\langle \Phi\rangle > 0.5$ and~$\langle \bar{\Phi}\rangle > 0.5$ for~$T/T_{\chi,0} \gtrsim 0.8$ and~$T/T_{\chi,0} \gtrsim 0.7$, respectively.
However, we observe a sudden steep\footnote{Within our numerical precision, we observe that
the change in~$\langle \Phi\rangle$ and~$\langle \bar{\Phi}\rangle$ is very steep at this point but sill continuous.} 
increase in the results for~$\langle \Phi\rangle$ and~$\langle \bar{\Phi}\rangle$ as a function of temperature
in this part of the phase diagram, e.g. at~$T/T_{\chi,0}\approx 0.43$ for~$\mu/m_q=1.02$ as illustrated by the 
black circles in Fig.~\ref{fig:ploops}. In the low-temperature regime, $\langle \Phi\rangle$ and~$\langle \bar{\Phi}\rangle$
are still small, suggesting that the inhomogeneous phase is well in the effectively confined phase. The steep increase can be traced back to the fact that the system undergoes
a second-order chiral phase transition from the inhomogeneous phase to the chirally symmetric phase at this point. 
Note that we only depict the phase diagram for the $\mu$-dependent parametrization in Fig.~\ref{fig:pd}.
We expect that this increase is smeared out
in the case of finite current quark masses when the chiral phase transition turns into a crossover as well.

The situation is different when we choose~$T_0$ to depend on~$\mu$. We then find that the deconfinement phase transition temperature
is shifted to lower temperatures. This is illustrated in Fig.~\ref{fig:ploops} where we show~$\langle \Phi\rangle$ and~$\langle \bar{\Phi}\rangle$ (red squares)
as a function of~$T/T_{\chi,0}$ for~$\mu/m_q=1.02$. Both~$\langle \Phi\rangle$ and~$\langle \bar{\Phi}\rangle$ now 
increase more rapidly as a function of temperature. 
Again, we observe a steep increase in these quantities at low temperatures which is driven by the chiral transition from the inhomogeneous
to the chirally symmetric phase at~$T=T_{\chi}(\mu)$. In fact, we find that
$\langle\Phi\rangle \approx 0.5$ at~$T=T_{\chi}(\mu)$, suggesting that the chiral phase transition temperature coincides 
approximately with the deconfinement phase transition temperature $T_\mathrm d$. The adjoint
Polyakov loop~$\langle \bar{\Phi}\rangle$ increases even stronger than~$\langle\Phi\rangle$ 
at low temperatures. In summary, when choosing~$T_0$ to be $\mu$-dependent, we do not observe
the emergence of a quarkyonic phase as found for~$T_0=\text{const.}$, even in the regime where inhomogeneous chiral condensation 
is found to be energetically favored.

\section{Conclusions}\label{sec:conc}

In the present work we have discussed the phase structure of the two-flavor PNJL model in the chiral limit as an effective 
low-energy model for QCD. 
We have particularly focussed on a discussion of the existence of inhomogeneous phases at low temperatures and 
large values of the chemical potential and on the question how these phases are affected by the confining dynamics being parametrized in terms
of the Polyakov loop potential. To this end, we have computed the effective potential with the fermion doubling trick~\cite{Braun:2014fga} and 
searched numerically for inhomogeneous ground states. We parametrized the latter by using a simple
single-cosine ansatz in the $\sigma$ channel with the amplitude and the frequency as variational parameters in the numerical
search for the ground state. This choice for the inhomogeneity
is motivated by the exact solution of the one-dimensional GN model 
where the inhomogeneity approaches this functional 
form close to the phase boundary~\cite{Thies:2003kk,Schnetz:2004vr,Thies:2006ti}. 
For higher dimensional NJL- and GN-type models, even the general functional form of the exact ground state is unknown. In this sense,
we only probe one specific form of the inhomogeneity. For our present study, however, we consider this to be sufficient as 
we primarily aimed at a qualitative 
understanding of how the emergence of inhomogeneous phases is affected by
the confining gauge dynamics. To this end, we expect that the knowledge of the exact
functional form of the inhomogeneity is not crucial. 
The strength of our present approach is that it allows to study different parametrizations and functional forms
of the Polyakov loop potentials at comparatively low computational cost. Here, we have used two prominent and
frequently used parametrizations
of this potential distinguished by a $\mu$-independent and a $\mu$-dependent choice for the parameter~$T_0$, respectively.

In accordance with earlier studies~\cite{McLerran:2007qj,Schaefer:2007pw}, where the possibility of inhomogeneous chiral condensation
has not been taken into account explicitly, we observe for constant~$T_0$ that a quarkyonic phase emerges between a low-temperature confined phase with broken chiral symmetry
and a high-temperature deconfined chirally symmetric phase, even when we now allow explicitly for inhomogeneous 
condensation in our studies, in accordance with Ref.~\cite{Buballa:2014tba}. Taking into account a possible $\mu$-dependence of~$T_0$, we observe that
the quarkyonic phase vanishes and we are only left with a low-temperature confined phase with spontaneously broken chiral symmetry
and a deconfined chirally symmetric phase, even in the regime of the phase diagram where inhomogeneous condensation is energetically favored.
Moreover, we find that the size of the inhomogeneous phase shrinks when we employ a $\mu$-dependent~$T_0$. 

Clearly, our present work only demonstrates that the dynamics at large chemical potential
is significantly affected by the parametrization of the gauge dynamics
in terms of the Polyakov loop potential. The present study is not completely conclusive 
with respect to the true phase structure of PNJL-type models (let alone QCD)
at large chemical potential. It only points to presently still existing uncertainties in our understanding of even the general phase structure of QCD in this regime.
Further studies with different functional forms of the Polyakov loop potential may be required to improve our understanding in this respect. In particular,
improved computations of this potential based on, e.g., functional RG approaches 
along the lines of Refs.~\cite{Braun:2007bx,Braun:2009gm,Haas:2013qwp} taking into account the backreaction of the 
quarks on the gluodynamics, may be necessary to further narrow down these uncertainties
and ultimately help to better the predictions for the equation of state at 
high densities.

{\it Acknowledgments.--~} The authors would like to thank J.~M.~Pawlowski for useful discussions.
JB, SR, and DR acknowledge support by HIC for FAIR within the LOEWE program of the State of Hesse. 
In addition, JB and SR acknowledge support by the Deutsche Forschungsgemeinschaft (DFG) through contract SFB 634.
JB and DR acknowledge support by the DFG under Grant BR \mbox{4005/2-1}.

%
\bibliography{bib_source}

\end{document}